\documentclass{emulateapj}

\shorttitle{The Rotating Nuclear Star Cluster in NGC~4244}
\shortauthors{Seth et al.}

\begin{document}

\slugcomment{Accepted by ApJ, July 15, 2008}

\title{The Rotating Nuclear Star Cluster in NGC~4244}

\author{Anil C. Seth\altaffilmark{1,2}, Robert D. Blum\altaffilmark{3}, Nate Bastian \altaffilmark{4}, Nelson Caldwell\altaffilmark{1}, Victor P. Debattista\altaffilmark{5}, Thomas H. Puzia\altaffilmark{6}}

\altaffiltext{1}{Harvard-Smithsonian Center for Astrophysics, 60 Garden Street Cambridge, MA 02138}
\altaffiltext{2}{Center for Astrophysics Fellow, {\tt aseth@cfa.harvard.edu}} 
\altaffiltext{3}{National Optical Astronomy Observatory, 950 North Cherry Avenue, Tucson, AZ, 85721}
\altaffiltext{4}{Institute of Astronomy, University of Cambridge}
\altaffiltext{5}{Centre For Astrophysics, University of Central Lancashire, Preston PR1 2HE, UK, {\it RCUK Fellow}}
\altaffiltext{6}{Herzberg Institute of Astrophysics, 5071 West Saanich Road, Victoria, BC V9E 2E7, Canada}

\begin{abstract}

We present observations of the nuclear star cluster in the nearby
edge-on spiral galaxy NGC~4244 using the Gemini Near-Infrared Integral
Field Spectrograph (NIFS) with laser guide star adaptive optics.  From
a previous study of edge-on galaxies, this nuclear star cluster was
found to be one of a sample of clusters that appear flattened along
the plane of their host galaxies disks.  Such clusters show evidence
for multiple morphological components, with younger/bluer disk
components and older/redder spheroidal components.  Our new
observations of NGC~4244 show clear rotation of 30~km\,s$^{-1}$ within
the central 10~pc (0.5$\arcsec$) of the cluster.  The central velocity
dispersion is found to be 28$\pm$2~km\,s$^{-1}$.  The multiple stellar
populations inferred from the optical colors and spectra are seen as
variations in the CO line strength in the NIFS spectra.  The rotation
is clearly detected even in the older, more spheroidal stellar
component.  We discuss evidence for similar structures and kinematics
in the nuclear star clusters of other galaxies including M33 and the
Milky Way.  Our observations support two possible formation
mechanisms: (1) episodic accretion of gas from the disk directly onto
the nuclear star cluster, or (2) episodic accretion of young star
clusters formed in the central part of the galaxy due to dynamical
friction.

\end{abstract}


\section{Introduction}

Prominent nuclear star clusters (NSCs) are found in $\sim$75\% of
late-type spiral galaxies \citep{boker02} and early-type galaxies
\citep[][]{cote06}.  These NSCs are intrinsically very luminous, with
typical $M_I \sim -12$, and sizes similar to globular clusters
\citep[$r_{eff} \sim 5$pc;][]{boker04a}.  Recently, multiple studies
\citep{ferrarese06,wehner06,rossa06} have shown that NSC masses are
coupled to the mass of their host galaxy, following a similar
$M$-$\sigma$ relation as massive black holes (MBHs) and extending this
relation to lower mass galaxies.  This result suggests that the
formation of both types of objects are linked to the formation of
their host galaxies.  Some galaxies host both NSCs and MBHs, but the
relationship between these two types of objects is unknown
\citep{seth08}.  Unlike black holes, the assembly history of NSCs can,
in principle, be inferred from studies of their stellar populations
and kinematics.  Thus NSCs provide important tools for studying the
history of galaxy centers and their connection to the galaxies as a
whole.

In spiral galaxies, integrated optical spectra of NSCs show that they
nearly always have multiple stellar populations
\citep{walcher06,rossa06}. More specifically, NSCs typically have a
young stellar population ($<$100~Myr) but are dominated in mass by
stars older than $>$1~Gyr.  Further support to this picture comes from
the observations of several distinct populations in the color-magnitude
diagram of the nucleus of the Sgr dwarf spheroidal
\citep[M54;][]{monaco05,siegel07}.

In \citet{seth06} (Paper~I), we found from a study of edge-on galaxies
that in some NSCs these multiple stellar populations appear to be
associated with different morphological components.  These clusters
are flattened along the major-axis of the galaxy and have color maps
and morphologies that suggest they consist of young disks and older
more spheroidal components.  

We present here follow-up observations of the NSC in NGC~4244, the
nearest galaxy in the Paper~I sample \citep[D=4.3~Mpc,
1$\arcsec$=21~pc;][]{seth05a}.  In Paper~I, we presented integrated
optical spectra of this cluster that indicate the presence of multiple
stellar populations.  Although the detailed star formation history is
not well-constrained, populations of 50-100~Myr combined with older
populations are required to provide a good fit to the spectra.
Redshifted emission identified with an HII region offset by 19~pc
along the major axis of the cluster gives a lower-limit to the mass of
$\sim$2.5$\times$10$^6$~M$_\odot$.  This mass measurement is
consistent with the masses derived from the stellar population fitting
of the spectrum as well.

In \S2 we describe the observations and data reduction.  Our results on
the kinematics, morphology and stellar populations of the cluster are
presented in \S3.  We discuss the results in the context of other
galaxies and formation models in \S4.

\section{Observations \& Data Reduction}

K-band spectra of the NGC~4244 nuclear star cluster were obtained
using the Near-Infrared Integral Field Spectrograph (NIFS) on Gemini
North, an image-slicing integral field unit spectrograph.  The
spectrograph was used in conjunction with the ALTAIR Laser Guide Star
system.  The cluster itself was used as a tip/tilt star, but is too
faint ($r$=16.18) to get the highest-Strehl performance.  The cluster
was observed in K-band on the nights of February 9 and April 29 2007.
In total 5$\times$900 second exposures were obtained on-source and
4$\times$900 sec exposures off-source (sky).  The airmass of the
observations ranged between 1.11 and 1.42, and an A0 star, HIP50459
was observed on both nights at similar airmass to be used as a
telluric calibrator.  The exposures all had similar image quality,
with the PSF FWHM varying by $<$30\% between the exposures.  We note
that H-band spectral observations of the NGC~4244 NSC were obtained as
well, but due to a number of observational problems and a
significantly lower resolution, these data are not discussed here.

The data were reduced using the Gemini v. 1.9 IRAF package, utilizing
pipelines based on the NIFSEXAMPLES scripts.  To obtain a clean
telluric absorption spectra from the HIP50459 observations, the
spectra were processed as described for the science images below,
after which 1-D spectra were extracted and the Br$\gamma$ absorption
line was fit and removed.  

\begin{figure}
\plotone{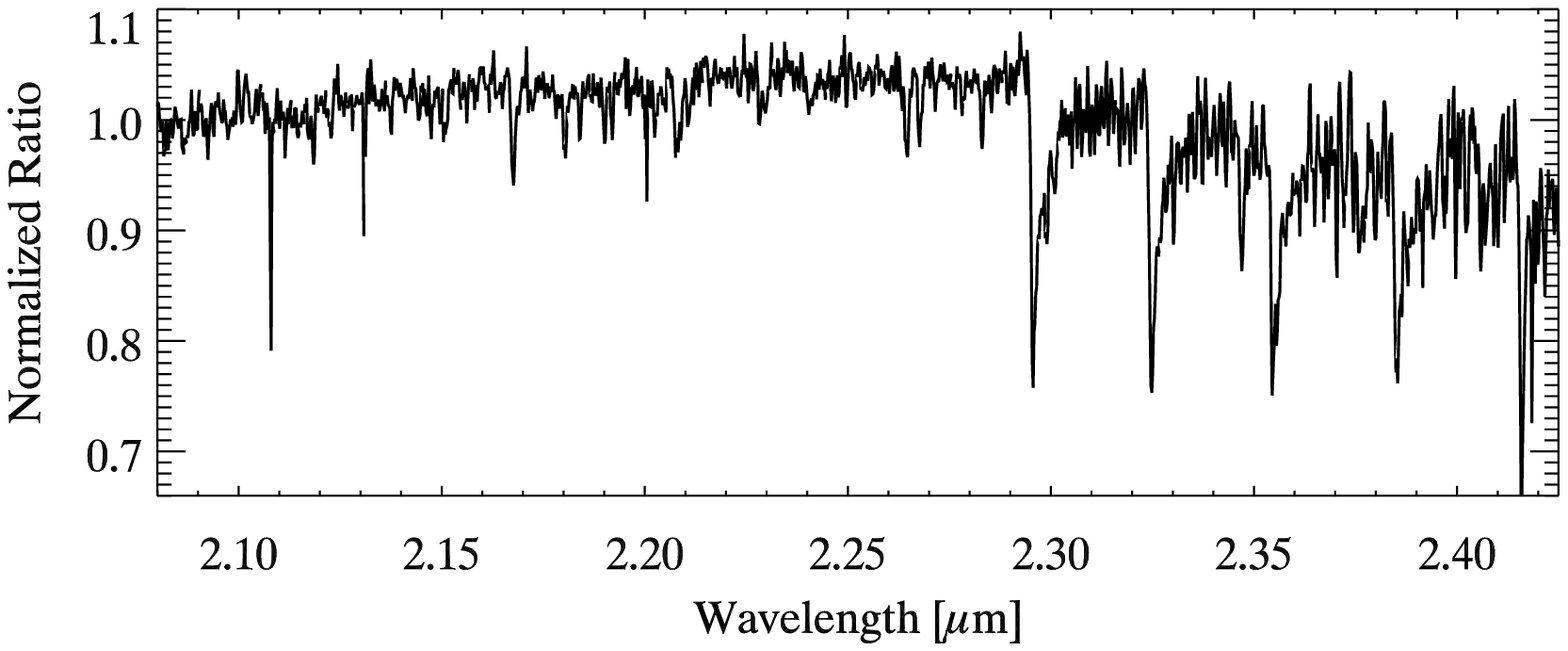}
\plotone{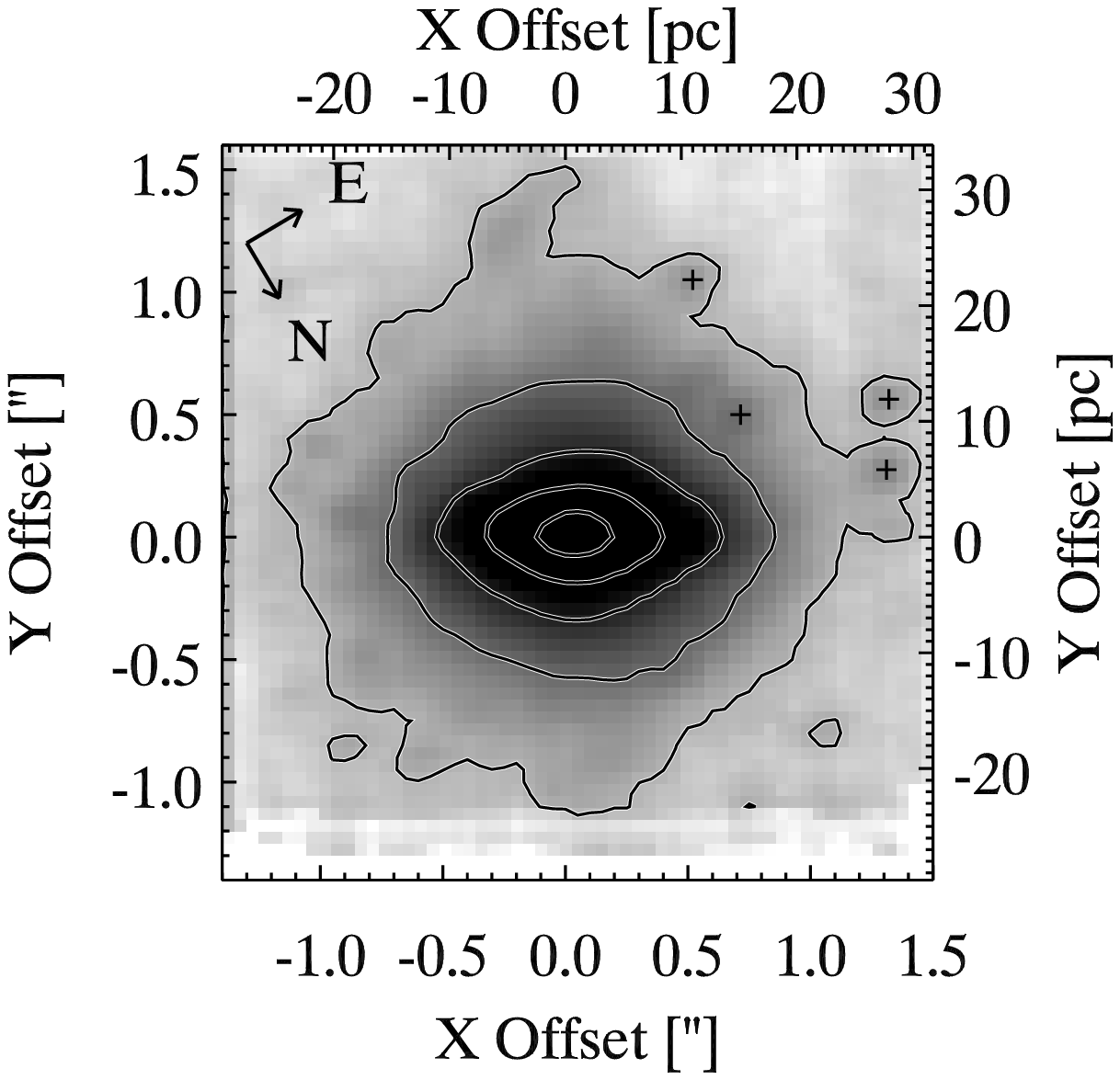}
\caption{{\it Top --} The spectrum within 0.1$\arcsec$ of the nuclear
star cluster center.  The plotted spectrum is the normalized flux
ratio of the galaxy spectrum to the Br$\gamma$-corrected A0V telluric
calibrator spectrum.  The CO bandheads dominate the red part of the
spectrum while Mg~I (21080,22830\AA), Br$\gamma$ (21676\AA), Na~I
(22075\AA), and Ca~I (22644-77\AA) lines are also seen. {\it Bottom
--} K-band image made from collapsing the final data cube along the
spectral direction.  Contours are at 16.5,15.5,14.5, 13.5, and 12.5
mag/arcsec$^2$.  Crosses show the four compact sources used to
determine the spatial resolution of the data.  The compass in the
upper left shows the rotation of the field (PA=147$^\circ$); the
x-axis runs roughly along the major axis of the galaxy.}
\label{firstfig}
\end{figure}

For the science exposures, an off-source exposure was subtracted from
each on-source exposure; for the February 9 data only one off-source
exposure was available and was thus subtracted from both images.  The
images were then sliced up and rectified based on Ronchi mask and arc
lamp images using NFTRANSFORM and spatially rebinned into image cubes
using NIFCUBE.  In this process, the original
0.043$\times$0.1$\arcsec$ pixels get rebinned into 0.05$\arcsec$
pixels which are block replicated in the long (slice) direction.  Each
data cube was then corrected for telluric absorption using the
NFTELLURIC program, which allows for interactive correction of the
small differences in the wavelength calibration and airmass between
the observations of the calibrator and each exposure.  The final data
cubes from each individual on-source exposure were combined using
GEMCOMBINE correcting for spatial and spectral offsets. The resulting
data cube has a $\sim$3$\arcsec$$\times$3$\arcsec$ field of view with
0.05$\arcsec$ spatial pixels and 2.132 \AA\ spectral pixels from
2.00-2.44 $\mu$m.  Based on measurements of the sky lines, the
spectral resolution is $\sim$4.1\AA\ (53~km\,s$^{-1}$) FWHM.  The
spatial resolution in the combined image was measured using four
compact sources after subtracting the NSC model (\S3.1) and found to
be 0.23$\arcsec$ FWHM.  These four compact sources are shown as
crosses in Fig.~\ref{firstfig}; three of the four are also detected as
point-like sources in the higher resolution (0.08$\arcsec$ FWHM) HST
images.  The spatial resolution of our data cube is roughly
$\sim$2$\times$ lower than that achieved when using a brighter
tip/tilt star (e.g. for our telluric calibrators).

A spectrum from the inner 0.1$\arcsec$ of the cluster is shown in
Fig.~\ref{firstfig}. The plotted spectrum is the normalized flux ratio
of the galaxy spectrum to the Br$\gamma$-corrected A0V telluric
calibrator spectrum.  All the spectral analysis presented below is
done using flux ratioed spectra.  In order to achieve a reasonable S/N
for analysis of the outer parts of the cluster, the data were
spatially binned using the Voronoi tesselation method described in
\citep{cappellari03}.  The S/N of each pixel was determined
empirically using a relatively featureless portion of the spectra
between 22140 and 22866\AA.  Individual pixels near the center of the
image have S/N$>$50 per pixel.  We binned pixels together in order to
obtain a S/N$\sim$25 in each bin, so near the center, the 'bins'
consisted of a single pixel.  We found that the S/N did not increase
as expected with binning (at least in part due to spectral rebinning
during the initial reduction), thus the outer bins have S/N of $<$10.

We also reconstructed a calibrated K-band image of the nuclear
star cluster for the morphological analysis presented in \S3.1.  To
do this, we first multiplied the spectra by a 9500K black body to
correct for the division by the A0 telluric calibrator, and then
multiplied this spectrum by the 2MASS $K_s$ filter response curve
before collapsing the data cube to an image.  We then calibrated the
zeropoint of this image by comparing the flux from the best fitting
model (described below) to the magnitude of the source in the 2MASS
Point Source Catalog \citep[$K_s$=13.55;][]{2mass_psc}.  Based on the
variation between models and the uncertainty in the sky-level on our
image, this zeropoint should be good to $\sim$10\%.  The image and
calibrated contours are shown in the bottom panel of
Fig.~\ref{firstfig}.

\section{Results}

In this section, we present results from our Gemini/NIFS K-band
observations.  We first discuss morphological fits to our high
resolution K-band image (\S3.1).  We then present the kinematics of
the cluster in \S3.2 and its stellar populations in \S3.3. The most
significant finding is that the NGC~4244 NSC is strongly rotating.
Comparison of the kinematics to the stellar populations and
morphological fits shows that this rotation is occuring even off the
major-axis of the cluster where older stellar populations appear to
dominate.

\subsection{Morphological Results}

We briefly describe our analysis of the morphology of the NGC~4244 NSC
using the high resolution K-band image of the cluster created from our
Gemini/NIFS data.  This information is used in \S3.3 to argue that the
entire cluster is rotating, and will be used as input for dynamical
modeling \citep{delorenzi08} using NMAGIC code \citep{delorenzi07}.

As we found for the optical data (see Paper~I, Fig.~4), a single
component elliptical King (or S\'ersic) profile fit to the K-band
image leaves behind significant residuals in the plane of the galaxy.
Combined with the optical color difference (Fig.~\ref{popfig} top
panel), this motivates us to fit the data using both a spheroidal
{\it and} disk component.  All models were convolved by a Gaussian PSF
with FWHM of 0.23$\arcsec$ during the fitting process.  We tried both
disk$+$King (as used in Paper~I) and a disk$+$S\'ersic profile.  Both
profiles provide fits with reduced $\chi^2$ values $\sim$4$\times$
smaller than the spheroidal component alone.  The values of the
derived parameters for these fits, and the disk$+$King F814W filter fit
from Paper~I, are shown in Table~\ref{fittable}.  The parameters for
the disk fits are the central surface brightnesss $\mu_0$, the scale
length $h_x$, and scale height $z_0$.  For the spheroidal component
fits, the parameters are the central surface brightness $\mu_0$ (King)
or effective surface brightness $\mu_{eff}$ (S\'ersic), effective
radius $r_{eff}$, axial ratio $(b/a)$, and S\'ersic index $n$.  The
derived K-band disk sizes are very similar to those derived at optical
wavelengths from the HST data despite the lower resolution of this
data.  The spheroid geometry is significantly more extended in the
K-band, but its size is quite sensitive to the fitted value of the
background.

Based on disk fits to the full galaxy from 2MASS presented in
\citet{seth05a}, the central luminosity surface density of the NSC is
$\sim$250$\times$ greater than that of the disk, while the luminosity
surface density at the edge of the NIFS field is comparable to the
level predicted by the galaxy disk model.  Deprojecting the disk
components, the luminosity volume density at the center of the NSC is
$\sim$10$^6\times$ that of the underlying galaxy disk.  The local
volume density of the cluster thus dominates the mass in the central
regions of the galaxy out to the edge of our data.

\begin{deluxetable}{lcccc}
\tablewidth{0pt} 
\tablecaption{Results of Morphological Fits \label{fittable}}
\startdata

\multicolumn{5}{c}{{\it NSC Disk Component Fits}}\\
\tableline
 & $\mu_0$ [L$_\odot$/pc$^2$] & $h_x$ [pc] & $z_0$ [pc] & \\
\tableline
K-band {\bf disk}$+$S\'ersic & 1.4$\times$10$^5$ & 3.39 & 1.19 & \\
K-band {\bf disk}$+$King     & 1.9$\times$10$^5$ & 2.72 & 1.38 & \\
F814W {\bf disk}$+$King      & 9.7$\times$10$^4$ & 2.73 & 1.42 & \\

\multicolumn{5}{c}{{\it NSC Spheroidal Component Fits}}\\
\tableline
& $\mu_0$ [L$_\odot$/pc$^2$] & $r_{eff}$ [pc] & $(b/a)$ & \\
\tableline
K-band disk$+${\bf King}     & 4.5$\times$10$^4$ & 13.6 & 0.79 & \\
F814W disk$+${\bf King}      & 5.3$\times$10$^4$ & 5.70 & 0.73 & \\
\tableline
& $\mu_{eff}$ [L$_\odot$/pc$^2$] & $r_{eff}$ [pc] & $(b/a)$ & $n$ \\
\tableline
K-band disk$+${\bf S\'ersic} & 8.7$\times$10$^3$ & 10.9 & 0.81 & 1.68 \\
\enddata
\tablecomments{All quantities derived assuming a distance of 4.3~Mpc.}
\end{deluxetable}

\vspace{0.3in}

\subsection{Kinematic Results}

We first describe the two ingredients used to derive the kinematics of
the NGC~4244 nuclear star cluster: (1) spectral templates, and (2) a
method to determine the line-of-sight velocity distribution (LOSVD)
from the binned spectra based on the templates.

Spectral templates were obtained from two sources.  Three template
stars were observed with NIFS by \citet{davidge08} in a configuration
identical to ours.  Spectra for these stars were kindly provided by
the author.  Their spectral types: K1III, K5III, and M0III are
representative of the spectral types of the red giant branch (RGB) and
asymptotic giant branch (AGB) stars most likely to dominate our
spectrum.  We also use a larger library of spectral templates that has
been observed by C. Winge using the somewhat higher resolution
(3.35~\AA\ FWHM) GNIRS spectrograph on Gemini
South\footnote{http://www.gemini.edu/sciops/instruments/nir/spectemp/index.html}.
This library contains stars from spectral type F7-M3 and includes
dwarf, giant and some supergiant stars.

To determine the LOSVD of each spectrum we used the penalized pixel
fitting (PPXF) method of \citet{cappellari04}.  The program finds the
best matching set of templates and convolves them to find the best
fitting LOSVD, which is parametrized using a Gauss-Hermite series.
The non-gaussian terms \citep[$h3$ and $h4$, see][]{vandermarel93} in
the LOSVD are penalized, thus the solution typically converges to a
Gaussian LOSVD at low S/N.  The method also compensates for any
low-frequency differences between the template and galaxy spectral
shape using Legendre polynomials.  We fit the area of the spectrum
from 22900\AA\ to 24000\AA, which is dominated by strong CO lines.  We
calculated errors on the LOSVD parameters using Monte Carlo tests that
consisted of adding noise to the spectrum based on the measured S/N
and remeasuring the LOSVD 250 times without penalizing non-Gaussian
distributions.  The typical standard deviation on the velocity and
dispersion for pixels with S/N$\sim$25 is $\sim$3~km\,s$^{-1}$.  When
using GNIRS templates, we convolve the template spectra by a Gaussian
with $\sigma=13$~km\,s$^{-1}$ to correct for the slightly higher
resolution of the GNIRS data.

\begin{figure}
\plotone{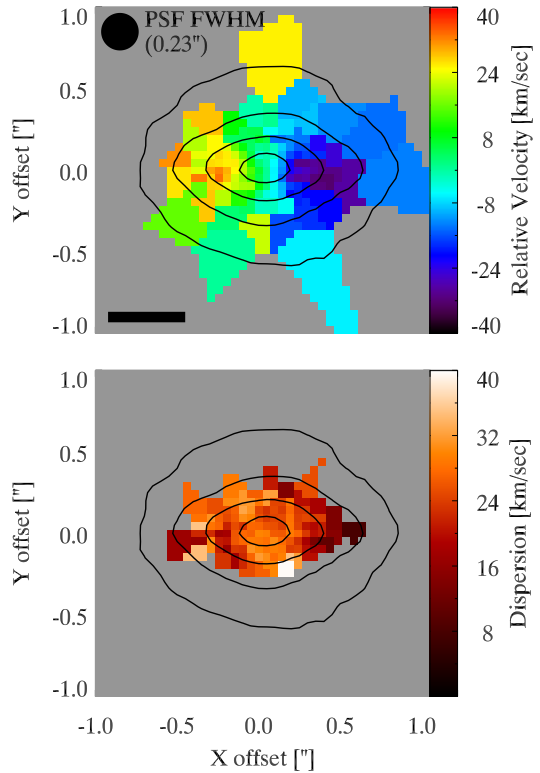}
\caption{{\it Top --} Color indicates the measured radial velocity for
the Voronoi binned data using templates observed with NIFS.  Gray
areas indicate spectra with S/N $<10$ and/or errors $>$ 15
km\,s$^{-1}$.  Rotation of $\sim$30~km\,s$^{-1}$ is clearly visible
along the major axis.  Contours show the K-band image at
15.5,14.5,13.5, and 12.5 mag/arcsec$^2$.  The black bar indicates
10~pc (0.47$\arcsec$). {\it Bottom --} Velocity dispersion
measurements with S/N $>$15 and errors $<$10~km\,s$^{-1}$. The
dispersion drops away from the center indicating a relatively cold
disk population.}
\label{velfig}
\end{figure}

We show the radial velocity map derived using PPXF on each spectral
bin in the top panel of Figure~\ref{velfig}.  We show only data with
reliable measurements by clipping bins with S/N$<$10 or errors on the
velocity above 15~km\,s$^{-1}$.  {\it The radial velocity map shows
strong rotation in the NGC~4244 NSC.}  The rotation velocities reach
an amplitude of 30~km\,s$^{-1}$ $\sim$0.4$\arcsec$ from the nucleus.
Assuming a Keplerian disk, this rotation suggests a mass of
1.7$\times$10$^6$ solar masses enclosed within the central 8~pc.
Beyond this radius, there is some suggestion the velocity may drop,
however due to the low S/N at large radii this result is not robust.

Based on the morphological results presented in the previous section,
the NSC dominates the mass in the center of NGC~4244, and thus is
likely to be kinematically distinct from the rotation of the overall
galaxy disk.  \citet{olling96} has presented a study of the HI
rotation curve of NGC~4244 with resolution as high as 10$\arcsec
\times$20$\arcsec$.  The HI gas rotates in the same direction as the
NGC~4244 NSC.  The innermost point in their derived rotation curve is
at 24$\arcsec$ where the galaxy rotation is 35~km\,s$^{-1}$,
increasing to 60~km\,s$^{-1}$ at 47$\arcsec$.  Given that we observe
rotation of $\sim$30~km\,s$^{-1}$ within the central half-arcsecond,
this suggests that the rotation of the NSC is distinct from the gas
component of the galaxy disk.  It would be interesting to compare the
NSC rotation to the stellar rotation in the galaxy disk, however, this
would require relatively deep long-slit spectroscopy
\citep[e.g.][]{yoachim08a} since the surface brightness of the galaxy
beyond the nuclear star cluster is quite low.


The line-of-sight velocity dispersion is a somewhat more difficult
quantity to measure accurately due to the low dispersion of the
cluster as a whole.  Our spectra have 2.1\AA/28~km\,s$^{-1}$ pixels
and 4.1\AA/53~km\,s$^{-1}$ resolution (FWHM), while the dispersions of
NSCs in late-type galaxies are typically $\sim$25~km\,s$^{-1}$
\citep{walcher05}.  Nonetheless, we are able to make an accurate
measurement of the central dispersion of the cluster, $\sigma =
28\pm2$~km\,s$^{-1}$.  Convolution of the templates with this
dispersion substantially (by a factor of 2) improves the $\chi^2$ of
their fit to the central spectrum of the cluster, giving a reduced
$\chi^2$ of 0.9 and 1.1 using the GNIRS and NIFS templates
respectively.  The bottom panel of Fig.~\ref{velfig} shows a map after
clipping data with S/N$<$15 or errors in the dispersion
$>$10~km\,s$^{-1}$.  The velocity dispersion drops along the major
axis to to values of 10-20~km\,s$^{-1}$ at precisely the area where
rotation is seen most strongly indicating that we are indeed seeing a
relatively cold rotating disk.  This dispersion is close to the limit
of what we can reliably recover from our relatively low resolution
observations, thus these values may represent upper limits.  It is
unclear whether the increased dispersion we see in the center is a
result of unresolved rotation or a genuinely hot component.

The use of the templates from NIFS and GNIRS give very similar results
for both the velocities and dispersions.  The GNIRS templates selected
for fitting the spectra are primarily at spectral types close to those
(K-M giants) represented by the 3 NIFS spectra.  For almost all the
spectra, our measured non-Gaussian $h3$ and $h4$ components of the
LOSVD were consistent with zero.  However, due to our low S/N and
resolution, we would not expect to be able to detect any non-Gaussian
signatures \citep{cappellari04}.

\subsection{Stellar Populations and Connection to Kinematics}

\begin{figure}
\plotone{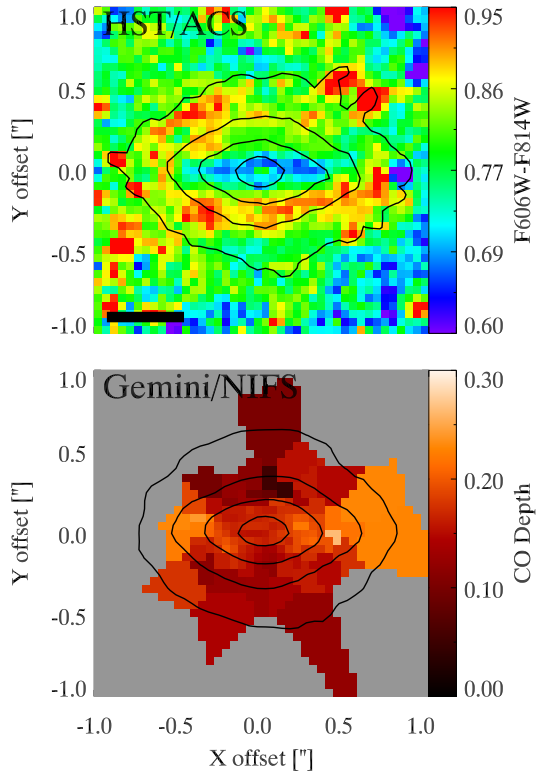}
\caption{2$\arcsec$$\times$2$\arcsec$ images of {\it top --} the F606W-F814W color
map from the HST/ACS data described in Paper~I. The image has been
rotated to match the orientation of the NIFS data.  The contours
indicate the F814W surface brightness and were chosen to roughly match
the K-band contours in \ref{velfig} and in the bottom panel.  The black
bar indicates 10~pc (0.47$\arcsec$). {\it Bottom --} the CO line strength
with larger values indicating deeper CO lines. The contours show
K-band contours as in \ref{velfig}.}
\label{popfig}
\end{figure}

From the optical data described in Paper~I, the NGC~4244 NSC has two
distinct stellar populations -- a bluer population dominates in the
midplane with a redder population dominating above and below the
plane.  This is clearly seen in the HST based color-map shown in the
top panel of Fig.~\ref{popfig}.  We would expect the NIFS data to
resolve these different populations despite the 2$\times$ lower
resolution.  If we make a map of the CO line strength (bottom panel of
Fig.~\ref{popfig}), the bluer areas of the HST image clearly have
stronger CO absorption than the redder areas above the midplane. The
CO line strength is determined by comparing the average depth of the
line at rest wavelengths of 22957$\pm$52\AA\ to the continuum measured
just to the blue of the CO bandhead \citep[$^{12}$CO~(2,
0);][]{kleinmann86}.  We also measured these values for the GNIRS
template stars for comparison.  The CO line strength values for the
NSC range from $\sim$0.10 above and below the major axis typical of
late G-type giant stars ($T_{eff} \sim 4800$~K), to $\sim$0.25 along
the major axis, typical of K3-K6 giants ($T_{eff} \sim 4100$~K).  We
caution that the CO depth depends on the gravity of the stars as well
as their effective temperature.  We note that while Mg, Ca, and
Br$\gamma$ absorption are clearly detected in the highest S/N spectra,
these features have lower S/N than the CO feature and do not show any
obvious variations across the NSC.  No emission features are detected. 

\begin{figure}
\plotone{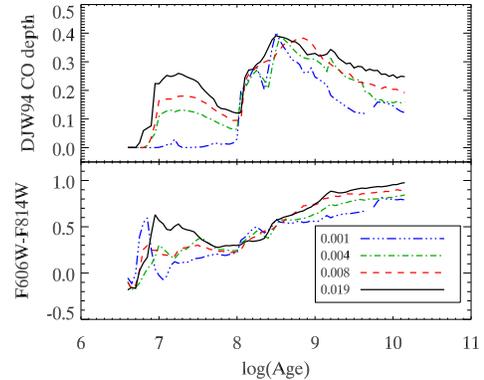}
\caption{Models of the optical color CO line strength using Padova
isochrones \citet{marigo08} at Z=0.004, 0.008, and 0.019.  The top
panel shows the CO line strength derived using relations from
\citet{doyon94}, while the bottom panel shows the F606W-F814W optical
color.}
\label{modelfig}
\end{figure}

To better understand the relation between the optical colors and CO
line strengths seen within the cluster, in Fig.~\ref{modelfig} we
model these quantities using Padova isochrones which include an
updated treatment of the asymptotic giant branch (AGB)
\citep{marigo08}.  We sampled 10$^7$ M$_\odot$ of stars from a
\citet{kroupa01} initial mass function and then obtained effective
temperature, gravities, and the F606W, F814W, and K-band magnitudes
for each star from the isochrones at ages from 4 Myr to 14 Gyr.  To
obtain estimates for the CO line strength, we used the relations from
\citet{doyon94}, who fit CO line strength as a function of effective
temperature for dwarfs, giants, and supergiants.  The total CO line
strength at each age was calculated as the mean CO line strength
weighted by the K-band luminosity of each star.  We note that the
broader \citet{doyon94} CO index has values that are 20-50\% lower
than the one we measure here, nonetheless, the trends of CO depth with
age should be robust.  We also caution that the models remain very
uncertain, especially given that the CO strength can be dominated by
the coolest, thermally pulsing AGB stars, which are very difficult to
model \citep{marigo08}.  The CO line strength shows peaks at two ages;
first at 10-20 Myr due to the presence of bright red supergiant stars,
and second at $\sim$300~Myr from AGB stars.  The CO evolution of these
models is qualitatively similar to that seen in the models of
\citet{maraston05}, which also include contributions from thermally
pulsing AGB stars, but differs from earlier models that don't include
this phase and thus are lacking the second peak in CO depth
\citep[e.g.][]{leitherer99}.

\begin{figure}
\plotone{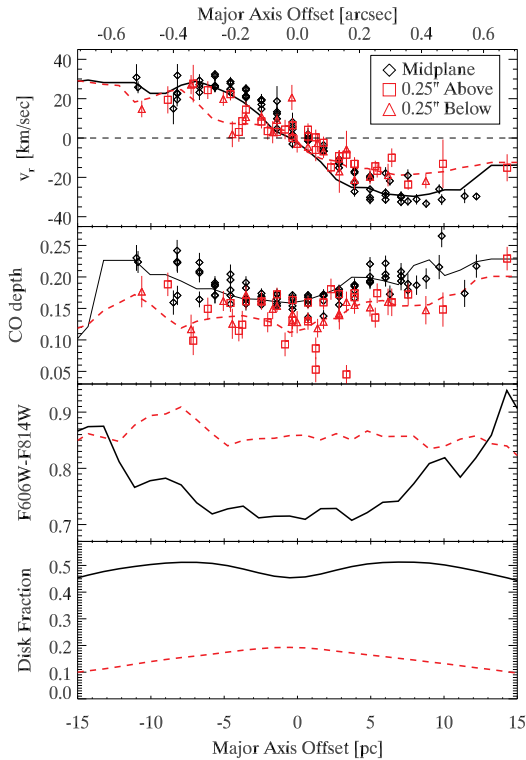}
\caption{Profiles of (from top-to-bottom) the relative radial
velocity, CO line strength, optical (F606W-F814W) color, and fraction
of light expected from the disk based on the morphological models.
Each quantity is plotted for the midplane (black solid line and
symbols) and for slices above and below (red dashed line and symbols)
the major axis.  Lines show vertically binned averages in each slice,
while the symbols in the top two panels show the values of each bin
that fell within our slices.}
\label{profilefig}
\end{figure}

We can use these models to qualitatively analyze the variations in
stellar population seen across the cluster in Figs.~\ref{popfig} and
\ref{profilefig}.  The extinction is not well known, however
spectroscopic fits suggest an $A_V \sim 0.5$ corresponding to a
reddening of $\sim$0.2 magnitudes in the F606W-F814W color.  For the
youngest stars, we expect a metallicity of Z$\sim$0.008 based on the
100~km\,s$^{-1}$ circular velocity of NGC~4244 \citep{garnett02}.
However, the combination of the red color and low CO values above and
below the midplane of the cluster suggest an old ($\gtrsim$1~Gyr),
more metal-poor population.  Along the major axis there is a slight
reddening in the optical color along with an increase in the CO line
strength.  One possible explanation for this is an age gradient within
the disk; for instance if the center had an age of $\sim$100~Myr
years then the redder colors and deeper CO could result from slightly
older or younger populations at 5-10~pc. We caution that the stellar
populations at any point in the cluster are clearly mixed and thus
information determined from comparing the observations to
single-stellar population models should be viewed with caution.  With
the maturing of NIR stellar models and spectral libraries
\citep[e.g.][]{lancon07,marmolqueralto08}, future population synthesis
studies may be able to provide more robust results on the resolved
star formation history from this type of data.  Lastly, we note that
it is possible for thermal emission from warm dust to reduce the
strength of the CO absorption.  However, the low star formation
activity in the NSC indicated by the lack of NIR emission lines
suggests that dust emission probably does not play a large role in
this cluster.

We now turn to the relation between the stellar populations and the
rotation.  Figure~\ref{profilefig} shows the derived velocity, CO line
strength, optical color, and disk fraction of the NGC~4244 NSC as
function of position in slices along, above, and below the major axis.
The slices are each 0.25$\arcsec$ in height, and the plotted lines are
the result of averaging the maps (i.e. Fig.~\ref{velfig}
and~\ref{popfig}) along the vertical direction in each slice at every
pixel along the x-direction.  The velocity and CO measurements were
made using the Voronoi binned data; the symbols in these panels show
the light-weighted positions of the bins coded by which slice they
fell in.

The top panel of Figure~\ref{profilefig} shows that the rotation is
not confined to the midplane, but is present $\sim$0.25$\arcsec$ above
and below the midplane as well.  The bottom three panels show that the
slices above and below the plane are clearly sampling a
different component/stellar population from the midplane.  The CO
lines are less deep away from the plane, while the optical color is
redder.  The disk fraction in the bottom panel is based on the results
of our morphological descomposition from \S3.1.  We calculated the
fraction of disk light along our slices from the best-fitting
S\'ersic$+$Disk model after convolving each component with the PSF.
For the area above and below the major axis, only 10-20\% of the light
in our NIFS observations appear to be coming from the disk.  Taken
together, {\it these plots show that the whole cluster is rotating,
not just the bluer, younger disk component.}  This finding places
strong constraints on the formation of this NSC as detailed in the
following section.

\section{Discussion \& Conclusions}

In NGC~4244, we find a flattened nuclear star cluster with multiple
components, a younger disk structure and an older spheroidal structure
(Paper~I).  From our NIFS data, we have shown that the cluster is
rotating at $\pm$30~km\,s$^{-1}$ within the central 10~pc of the cluster.
This rotation is not confined just to the younger disk component, but
also appears to be present in the older, more spheroidal component. 

There are two basic models for NSC formation: (1) globular clusters
which collect at the center due to dynamical friction
\citep{tremaine75,lotz01}, or (2) episodic accretion of gas onto the
center followed by {\it in situ} star formation
\citep{milosavljevic04,bekki07}.  A third alternative (3), combining
these two scenarios, has recently been proposed by Milosavljevic \&
Agarwal (2008, {\it private communication}) in which NSCs form from
young star clusters in the inner part of the galaxy disk which are
massive enough to accrete onto the nucleus.  These mechanisms are not
exclusive; NSCs could form from a combination of these effects, and
different types of galaxies may have different dominant mechanisms.
However, our observations for NGC~4244 strongly conflict with the
first scenario.  If a significant fraction of the NSC were formed from
globular clusters distributed in a spheroidal distribution no rotation
would be expected.  Thus the primary formation mechanism must be
through episodic accretion of material from the disk of the galaxy,
either gas or young star clusters.  Distinguishing between gas and
star cluster accretion may be possible through comparison of our
kinematic data with simulations.  Direct evidence of gas accretion
into the central $\sim$10 parsecs is seen in a couple nearby nuclear
star clusters using CO observations \citep{schinnerer03,schinnerer06}.

There is substantial evidence that the structure we see in NGC~4244 is
not unique, and in fact may be typical for NSCs in spiral galaxies.
In Paper~I, 5/9 NSCs in edge-on spirals were significantly flattened
along the major axis of the galaxy disk, with three showing clear
composite morphologies.  NGC~4244 is the nearest of these galaxies and
thus presents the clearest view of these structures.  There is also
evidence for similar structures in earlier type galaxies;
\citet{balcells07} find a number of disky ``nuclear extended
components'' with scale lengths as small as 5 pc in a sample of
early-type spirals and elliptical galaxies.  Previous studies of
nuclei in large samples of spirals \citep{boker02,carollo02} have
focused on face-on galaxies making the structures in the disk plane
difficult to detect.  And very few galaxies have the pc-scale quality
of kinematic data that we have here.  However, where this data exists
there is evidence for nuclear structures similar to those presented
here.

The Milky Way hosts a nuclear star cluster centered on Sgr~A$^*$, first
noted by \citet{becklin68}.  The luminosity profile of the inner
$\sim$100~pc of the Milky Way is distinct from the underlying bulge
and disk profiles \citep[e.g.][]{serabyn96}. This component is clearly
flattened along the galactic plane \citep[axial ratio of
0.45;][]{catchpole90}, but less so in the central 30 pc, which led
\citet{launhardt02} to propose the existence of two components, a
3$\times$10$^7$ Msol NSC dominating the central 30~pc
(with $r_{eff} \sim 10$~pc) and a more massive nuclear stellar disk
extending out to 100-200~pc.  Numerous young OH/IR stars are found,
primarily within the central 40~pc and these have a flattened
distribution with rotation of $>$100~km\,s$^{-1}$ \citep{lindqvist92}.
Based on integrated K-band spectra, \citet{mcginn89} find that the
central dispersion is $\sim$125~km\,s$^{-1}$ and drops quickly with
increasing radius, while the rotation increases outwards equalling the
rotation at a radius of $\sim$3~pc, the edge of their study.  Compared
to the NGC~4244 NSC, the Milky Way appears quite similar, but scaled
up by $\sim$10$\times$ in mass and $\sim$3$\times$ in size, rotation,
and dispersion as would be expected for a more massive galaxy.  The
very center ($<$1~pc) of the MW NSC has been intensively studied and
shows complex dynamics with striking differences between older stars
and very young ($<$10 Myr) stars. The latter show ordered rotation
(and counter rotation!) and provide evidence for processes 2
\citep{genzel03} and 3 \citep[e.g.][]{maillard04,lu05} described
above.

There is also evidence for flattening and rotation in the M33 NSC.
This cluster is elongated along the major axis of the galaxy with an
axial ratio of $\sim$0.85 \citep{lauer98,matthews01}.  At a radius of
$\sim$1$\arcsec$ (4~pc), the rotation is $\sim$8~km\,s$^{-1}$ and
dispersion is 27~km\,s$^{-1}$ \citep{gebhardt01}.  Even considering
that the orientation is likely not edge-on, it appears that the
rotation does not dominate in the M33 NSC to the same extent as in
NGC~4244.

In conclusion, we have shown above that the nuclear star cluster in
NGC~4244 is rotating.  The rotation is seen in both the young disk and
older spheroid components, suggesting that accretion of gas or star
clusters from the disk is the dominant mechanism in the formation of
this NSC.  While there is some evidence that other NSCs in spiral
galaxies have similar structures (and thus formation histories), a
better understanding of NSC formation across Hubble type will require
a detailed study resolving the stellar populations and kinematics
of numerous nearby NSCs.

Acknowledgments: The authors thank Tracy Beck, Andre Wong, and Richard
McDermid for their help with the observation planning and analysis.
We thank Tim Davidge for giving us his spectral templates.  This paper
was much improved by the suggestions of the referee, Torsten B\"oker.
We also acknowledge helpful conversations with Thomas Puzia, Kevin
Covey, Dawn Erb, and Andr\'es Jord\'an.  Anil Seth gratefully
acknowledges the support of the CfA Postdoctoral Fellowship.  Based on
observations obtained at the Gemini Observatory, which is operated by
the Association of Universities for Research in Astronomy (AURA),
Inc., under a cooperative agreement with the National Science
Foundation (NSF) on behalf of the Gemini partnership: the NSF (United
States), the Science and Technology Facilities Council (United
Kingdom), the National Research Council (Canada), CONICYT (Chile), the
Australian Research Council (Australia), the Minist\'erio da
Ci\'encia e Tecnologia (Brazil), and SECYT (Argentina).

{\it Facilities:} \facility{Gemini:Gillett (NIFS/ALTAIR)}, \facility{HST (ACS/WFC)}


\end{document}